\documentclass[aps,showpacs]{revtex4}
\usepackage{graphicx}
\usepackage{color}
\usepackage[T1]{fontenc}
\newtheorem{theorem}{Theorem}
\RequirePackage{amsfonts,amssymb,amsmath}

\begin{document}
\title{Quantum Singularities in  Ho\v rava-Lifshitz Cosmology}
\author{João Paulo M. Pitelli} 
\email{e-mail: pitelli@ime.unicamp.br} 
\author{Alberto Saa}
\email{e-mail: asaa@ime.unicamp.br}
\affiliation{
Departamento de Matem\'atica Aplicada,
Universidade Estadual de Campinas,
13081-970 Campinas,  SP, Brazil}

\begin{abstract}

The recently proposed Ho\v rava-Lifshitz (HL) theory of gravity is analyzed from the 
quantum cosmology 
point of view. By employing  usual quantum cosmology techniques,  we study the quantum Friedmann-Lema\^\i tre-Robertson-Walker (FLRW)   universe filled with radiation in the
context of HL gravity. 
We find that this universe is quantum mechanically nonsingular
in two different ways: the expectation value of the scale factor $\left<a\right>(t)$ never vanishes and, if we abandon the detailed balance condition suggested by Ho\v rava, the quantum dynamics of the universe is uniquely determined by the initial wave packet and no boundary condition at $a=0$ is indeed necessary.
\end{abstract}


\maketitle

\section{Introduction}
In 2009, Ho\v rava proposed a new theory of gravitation \cite{horava} based on an anisotropic scaling of space $\bf x$ and time $t$ coordinates. The resulting theory, since then dubbed 
Ho\v rava-Lifshitz (HL) gravity,
 has proved to be power countable renormalizable.
One of its key points is that, even though 
 it does not exhibit relativistic invariance at short distances,  General Relativity (GR) is indeed recovered for low-energy limits. Some interesting consequences of this theory include
the existence of nonsingular bouncing universes \cite{wang,calgani,brandenberger} and the possibility that it may represent an alternative to inflation, since it might solve the flatness and horizon problem and generate scale invariant perturbations for the early universe without the need of exponential expansion \cite{wang2,kiritsis,mukohyama}.

Due to the asymmetry of space and time in the HL gravity, its natural framework is the Arnowitt-Deser-Misner (ADM) formalism \cite{adm}, where the spacetime metric $g_{\mu\nu}(t,{\bf x})$ is decomposed as usual 
 in terms of the 3-dimensional metric $h_{ij}(t,{\bf x})$ of the spatial slices 
 of constant $t$, the lapse function $N(t,{\bf x})$, and the shift vector $N^i(t,{\bf x})$. In his original work, Ho\v rava made an important assumption on the lapse function to simplify the HL gravitational action, the so-called ``projectablity condition'', namely  $N\equiv N(t)$. There are, nevertheless, extended models where this condition is relaxed. Projectable theories  give rise to a unique integrated Hamiltonian constraint,
leading to great complications when compared with the GR. The so-called 
 non-projectable theories, on the other hand, typically give rise to a local Hamiltonian constraint, as in GR. Healthy non-projectable extensions of the original HL gravitational action are discussed in details in case\cite{blas}. Fortunately, 
 since FLRW spacetimes are homogeneous and isotropic, the spatial integral can be dropped from the integrated Hamiltonian constraint in our case \cite{sotiriou,Bellorin:2011ff}, 
 yielding  a true local constraint even for the projectable case. For our purposes here,
 it suffices to
 consider  the simplest HL projectable theory in a FLRW spacetime. We notice that a
modified $F(R)$  HL theory in a FLRW spacetime has been recently considered in \cite{Elizalde:2010ep}, leading to very interesting results regarding the possible
unification between primordial inflation and dark energy. 
  Another important assumption originally  introduced  by Horava is the principle of ``detailed balance''. This condition, which states that the potential in the gravitation action follows from the gradient flow generated by a 3-dimensional action, reduces the number of independent coupling constants. Recently, it has became clear  that the detailed balance condition can be also relaxed \cite{sotiriou,wang3,Carloni:2009jc,Carloni:2010ji,Zhu:2011yu}. In particular, 
 in \cite{Carloni:2009jc,Carloni:2010ji}, 
the dynamical role and the consequences for the matter couplings of the  detailed balance condition in classical cosmology are detailed. 
  In this paper we will abandon  the detailed balance condition, since, as we will show, it gives rise to the most interesting quantum universes.
The limit where this condition can be recovered will also be discussed.

There have been  many attempts to incorporate Quantum Mechanics into GR. One of the first ones
was Quantum Cosmology. In Quantum Cosmology, we work with the Hamiltonian (ADM) formulation of GR, use Dirac's algorithm \cite{dirac} of quantization, {\em i.e.}, the substitution $\pi_q\to-i\delta/\delta q$, where $\pi_q$ is the canonical momentum associated with the  variable $q$ (which can be one of the three canonical variables in GR, $h_{ij}$, $N$ or $N^i$) and the imposition that the first-class constraints of the theory should annihilate the wave function of the spacetime. GR has four constraints, three of them just tell us that the wave function of the spacetime depends only on the intrinsic geometry of the spatial slices in the ADM decomposition, while the last one is a dynamical constraint which gives the dynamical equation of quantum cosmology, the so-called Wheeler-DeWitt equation \cite{dewitt}. The wave function is a priori defined on the space of all 3-metrics, called superspace, which 
 are in general very intricate infinite dimensional spaces. However, we can take advantage
of the symmetries of a homogeneous universe to freeze out  all but a finite number of degrees of freedom of the the metric and then quantize the remaining ones. These models are known as minisuperspace models.
Quantum cosmology in FLRW minisuperspace filled with a perfect fluid has been shown to be
viable and interesting  in the sense that the initial big-bang singularity is not present in such model since $\left<a\right>(t)\neq 0$ for all times  and the classical behavior of the universe is recovered for large times \cite{lemos,alvarenga}. Moreover, in this class of models, a certain evolution parameter of the fluid gives us a measure of time and one can investigate the evolution of the scale factor as the fluid evolves.

For static spacetimes, Horowitz and Marolf \cite{horowitz} found an original way of classifying a spacetime as quantum mechanically nonsingular. In their work, a spacetime is said to be quantum mechanically nonsingular if the evolution of quantum particles in the classical background is uniquely determined by the initial wave packet, {\em i.e.}, no boundary conditions at the classical singular points are necessary. This is equivalent to say that the spatial part of the wave equation is essentially a self-adjoint operator, {\em i.e.}, it has a unique self-adjoint extension (for a review about the mathematical framework necessary to define quantum singularities in static spacetimes, see \cite{pitelli}). In the GR context, the quantization of the FLRW minisuperspace filled with a perfect fluid does require a boundary condition at $a=0$ in order to assure the self-adjointness  the Wheeler-DeWitt equation, which, on the other hand, is necessary to guarantee a unitary time evolution.
Mathematically, the Hamiltonian operator corresponding to the evolution equation of the universe is not essentially self-adjoint. In this way the quantum dynamics of the universe is not unique since we do not know, in principle, which boundary condition we must apply at the initial singularity. However, as we will see later, it is possible to find quantum cosmologies in the HL gravity context for which the quantum evolution of the universe is unique, and no boundary condition for the wave function is indeed necessary.

In this paper we will apply the machinery of quantum cosmology to the HL theory of gravity. In particular, we will investigate the necessity of initial boundary conditions for the Wheeler-DeWitt equation and also the behavior of the universe, by  examining  the time evolution of the expectation value of the scale factor. A certain evolution parameter of the radiation filling the universe will play the role of the time coordinate.
 The   content of the universe will be introduced in the gravitational action via the Schutz formalism \cite{schutz1,schutz2},   demanding the recovering of the usual GR formulation in the low-energy\cite{saridakis}. The paper is organized as follows. Sections II, III, and IV present brief reviews of the results we need, respectively, the main definitions about quantum singularities, the HL theory of gravity, and the usual quantum cosmology in the GR context.
 Our main results are presented in the sections V and VI. The last section is devoted to some concluding remarks.

\section{Quantum Singularities}

Typical solutions of the Einstein field equations are known to exhibit singularities. They can be classified as \cite{konkowski}: quasiregular singularities, where the observer feels no physical quantity diverging, except at the moment when its worldline reaches the singularity (for instance,  the conical singularity of a cosmic string); scalar curvature singularities, where every observer approaching the singularity experience diverging tidal forces (for example, the singularity inside a black hole and, more important in the present context, the big bang singularity in FLRW cosmology); non scalar singularities, where there are some curves in which the observers experience unbounded tidal forces (for example, whimper cosmologies).
It is well known \cite{hawking} that under very reasonable conditions (the energy conditions), which basically state that gravity must be attractive, singularities are inevitable in GR. In this way, cosmological models with non-exotic fluids, as radiation or dust, present typically an initial singularity, known as the big bang singularity. Since we cannot escape this fact in GR, we hope that the quantum theory of gravitation will solve this issue, guiding us on how to deal with the singularities, or even excluding them at all. Unfortunately, we do not have such theory yet. However, there are several evidences that this theory would actually solve this problem. These evidences come with the introduction of quantum mechanics in GR in many different ways. In this paper we will highlight two distinct approaches.

The first approach is quantum field theory in curved spacetimes. In this framework, we analyze the behavior of quantum particles (or fields) in a classical curved background, which we assume to be a regular solution of the Einstein field equations. We adopt   Horowitz and Marolf's definitions\cite{horowitz} . In their work, they analyze the behavior of a scalar particle in singular static spacetimes possessing a timelike Killing vector field $\xi^\mu$. In such spacetimes, the wave equation can be separated into
\begin{equation}
\frac{\partial^2\Psi}{\partial t^2}=-A\Psi,
\end{equation}
where $A=-VD^i(VD_i)+V^2M^2$ and $V=-\xi^{\mu}\xi_{\mu}$, with $D_i$ being the spatial covariant derivative in a static slice $\Sigma$ not containing the singularity.  In principle, the domain $\mathcal{D}(A)$ of the operator $A$ is not known, so we choose as a first attempt $\mathcal{D}(A)=C_{0}^{\infty}(\Sigma)$. In this way, our operator is symmetric and positive definite. Howerver, this domain is unnecessarily  small, or in other words, the conditions on the functions are so restrictive that the operator $A$ is not self-adjoint. Its adjoint operator $A^{\ast}$ has a much larger domain $\mathcal{D(A^{\ast})}=\left\{\Psi\in L^2(\Sigma):A\Psi\in L^2(\Sigma)\right\}$. It is important to notice that we have chosen $L^2(\Sigma)$ as the Hilbert space of our quantum theory (for a discussion about this point see \cite{pitelli}). We must relax the conditions on the allowed functions in order to extend the domain of $A$ in such a way that $\mathcal{D}(A^{\ast})\to \mathcal{D}(A)$. If the extended  operator is unique, $A$ is said essentially self adjoint and its extension is given by $(\overline{A},\mathcal{D}(\overline{A}))$, where $\overline{A}$ is the closure of $A$ (for more detais see \cite{reed}). The time evolution of the particle will be then given by
\begin{equation}
\Psi(t)=\exp{\left(-i t \overline{A}^{1/2}\right)}\Psi(0),
\end{equation}
and the spacetime is said to be quantum mechanically nonsingular. However, if the extension is not unique, {\em i.e.}, if there exists infinitely many extensions $A_{\alpha}$, with $\alpha$ being a parameter such that to each $\alpha$ there corresponds one boundary condition at the singular point, then we have a different time evolution
\begin{equation}
\Psi_\alpha(t)=\exp{\left(-i t A_\alpha^{1/2}\right)}\Psi(0)
\end{equation}
for each $\alpha$. In this case, the spacetime is said to be quantum mechanically singular.
Similarly  to the classical case, when a spacetime is quantum mechanically singular, an extra information (a boundary condition) must be given in order to obtain the time evolution. In GR, in particular, we do need to tell what happen to the particle when it reaches the singularity.

The second approach we exploit here is quantum cosmology in minisuperspaces. In this framework,
 we consider a few degrees of freedom of the system (the rest is assumed to be frozen) and quantize the constraints of the theory via Dirac's algorithm. We impose $\left[a,p_a\right]=i$ (in units where $\hbar=1$), where $a$ is the scale factor of FLRW models, and  $\left[T,p_T\right]=i$, where $T$ is a parameter associated with the evolution of the fluid filling the universe, obtaining, in this way,  the Wheeler-DeWitt equation of the universe, which, as we will see,  is a Sch\"odinger like equation, from where we can define an internal product between two solutions and, therefore,   evaluate expectation values of observables.
 In this context, we define the universe as nonsingular if $\left<a\right>(t)\neq 0$ for all times. Since the operator $\hat{a}$ is positive in $L^2(0,\infty)$, we will have $\left<a\right>(t) = 0$ if the wave function representing the universe is sharply peaked at $a=0$. Note that this criterion is different from that one originally stated by DeWitt, which says that the universe is quantum mechanically nonsingular if $\Psi(a=0,t)\neq 0$ $\forall$ $t$. In fact, it was shown that this criterion is not enough to prevent singularities in quantum cosmological models \cite{lemos2}.

The two classifications of quantum singularities described above belong to completely different frameworks, but   we can apply the mathematical machinery used in static spacetimes in order
to  decide if the evolution of a wave packet governed by the Wheeler-Dewitt equation is unique
in a given quantum cosmology scenario.

\section{HL gravity}

In order to introduce the HL theory of gravity, let us 
first introduce the decomposition of the metric in the ADM form 
\begin{equation}
ds^2=-N^2c^2dt^2+h_{ij}(dx^i-N^idt)(dx^j-N^jdt),
\end{equation}
and then let us postulate that the dimensions of space and time are (in units of momentum) $[dx^i]=-1$ and $[dt]=-3$. This assumption assures that theory is power-countable renormalizable in four dimensions. In these units, we have
$[N]=[h_{ij}]=0$,
while $[N^i]=2$, leading to $[ds^2]=-2$.
Notice that the volume element, defined by
\begin{equation}
dV_4=N\sqrt{h}d^3{\bf x}dt,
\end{equation}
has dimension $[dV_4]=-6$.

The extrinsic curvature tensor, which measures how the spatial slices in the ADM decomposition of spacetime curves with respect to external observers, is defined by
\begin{equation}
K_{ij}=\frac{1}{2N}\left(\frac{\partial h_{ij}}{\partial t}-\nabla_{\left(i\right.}N_{\left.j\right)}\right).
\end{equation}
It is easy to see that it has dimension $[K_{ij}]=3$. The most general term 
 involving the extrinsic curvature tensor which is invariant under the group of diffeomorphism  of the spatial slices will define the kinetic term in the action. This term depends on two coupling constants $\alpha$ and $\lambda$ and is given by
\begin{equation}
S_K=\alpha\int{dtd^3{\bf x}\sqrt{h}N\left(K_{ij}K^{ij}-\lambda K^2\right)}.
\end{equation}
Note that $[\alpha]=0$, {\em i.e.}, $\alpha$ is a dimensionless constant. This is the reason why we made the choice $[dt]=-3$. 

The potential term for the gravitational action is given by
\begin{equation}
S_V=-\int{dtd^3{\bf x}\sqrt{h}NV[h_{ij}]},
\end{equation}
where $V[h_{ij}]$ is built out of the spatial metric and its spatial derivatives. Since $[dV_4]=-6$, we must have $[V[h_{ij}]]=6$ in order to assure that $S_V$ be a scalar. The most general action (without the detailed balance condition)  containing terms with dimensions less or equal than $6$ is given by (for more details see \cite{sotiriou})
\begin{equation}
S_{HL}=S_K+S_V,
\end{equation}
where
\begin{equation}\begin{aligned}
V[h_{ij}]&=g_0\zeta^6+g_1\zeta^4 R+g_2\zeta^2 R^2+g_3\zeta^2R_{ij}R^{ij}\\
&+g_4R^3+g_5R(R_{ij}R^{ij})+g_6R^{i}_{\phantom{i}j}R^{j}_{\phantom{j}k}R^{k}_{\phantom{k}i}\\&+g_7 R\nabla^2R+g_8\nabla_iR_{jk}\nabla^iR^{jk}.
\end{aligned}\end{equation}
Here the constant $\zeta$ has dimension $[\zeta]=1$ and ensures that all the coupling $g_a$ are dimensionless. In order to restore the units where $c=1$, {\em i.e.}, $[dx]=[dt]$ we need to
perform the transformation $dt\to\zeta^{-2}dt$.

Since $[R^i_{jkl}]=2$, as we go to lower momenta, the dominant action is
\begin{equation}
S_{IR}=\int{dtd^3{\bf x}N\sqrt{h}\left[\alpha\left(K_{ij}K^{ij}-\lambda K^2\right)-g_1 \zeta^4 R-g_0\zeta^6\right]}.
\end{equation}
We can now re-scale time and space so that $\alpha=1$ and $g_1=-1$, and set $c=\lambda=1$, leading to 
\begin{equation}
S_{IR}=\zeta^2\int{dtd^3{\bf x}N\sqrt{h}\left[\left(K_{ij}K^{ij}- K^2\right)+ R-g_0\zeta^2\right]}.
\end{equation}
Note that, by choosing
\begin{equation}
\zeta^2\equiv\frac{1}{16\pi G}, \;\;\; \Lambda=\frac{g_0\zeta^2}{2},
\end{equation}
we have the usual Einstein-Hilbert action
\begin{equation}\begin{aligned}
S_{GR}&=\frac{1}{16\pi G}\int{dtd^3{\bf x}N\sqrt{h}\left(K_{ij}K^{ij}- K^2+ R-2\Lambda\right)}\\
&=\frac{1}{16\pi G}\int{d^4x\sqrt{-^{(4)}g}\left(^{(4)}R-2\Lambda\right)},
\end{aligned}\end{equation}
where $^{(4)}g_{\mu\nu}$ and $^{(4)}R$ are the spacetime metric and Ricci scalar, respectively.

The full HL action we will consider hereafter  is
\begin{equation}\begin{aligned}
S_{HL}&=\frac{M_P^2}{2}\int dt d^3{\bf x}N\sqrt{h}\left(K_{ij}K^{ij}-\lambda K^2+R-2\Lambda\right.\\
&-g2M_P^{-2}R^2-g_3M_P^{-2}R_{ij}R^{ij}-g_4M_P^{-4}R^3\\&-g_5M_{P}^{-4}R(R_{ij}R^{ij})-g_6M_P^{-4}R^{i}_{\phantom{i}j}R^{j}_{\phantom{j}k}R^{k}_{\phantom{k}i}\\&\left.-g_7M_P^{-4}R\nabla^2R-g_8M_P^{-4}\nabla_iR_{jk}\nabla^iR^{jk}\right),
\end{aligned}\label{HL action}\end{equation}
where $M_P=1/\sqrt{8\pi G}$ stands for the Planck mass in $c=1$, $\hbar=1$ units.

\section{Quantum Cosmology in GR}

In the so-called Schutz formalism \cite{schutz1,schutz2} for the matter content of GR, the four-velocity of a perfect fluid is expressed in terms of six potentials in the form
\begin{equation}
U_{\nu}=\mu^{-1}(\phi_{,\nu}+\alpha\beta_{,\nu}+\theta S_{,\nu}),
\label{four-velocity}
\end{equation}
where $\mu$ and $S$ are, respectively, the  specific enthalpy  and the specific entropy of the fluid. The potentials $\alpha$ and $\beta$ are connected with rotations and, hence, they are not present in the FLRW universe due to its symmetry. The potentials $\phi$ and $\theta$ have no clear physical meaning. With the usual normalization 
\begin{equation}
U^{\nu}U_{\nu}=-1,
\end{equation}
Schutz showed that the action for the fluid in GR is given by
\begin{equation}
S_{f}=\int{d^4x\sqrt{-g}p},
\end{equation}
where $p$ is the pressure of the fluid, which is related  to the density by the equation of state $p=w\rho$. In this way, the total action for the spacetime filled with a perfect fluid is given by
\begin{equation}
S=\frac{M_P^2}{2}\int{d^4x\sqrt{-g}\left(R-2\Lambda\right)}+\int{d^4x\sqrt{-g}p}.
\end{equation}
Varying the above action with respect to the metric we get the usual Einstein equations
\begin{equation}
G_{\mu\nu}+\Lambda g_{\mu\nu}=M_P^{-2}T_{\mu\nu}, 
\end{equation}
with $T_{\mu\nu}$ given by
\begin{equation}
T_{\mu\nu}=(\rho+p)U_{\mu}U_{\nu}+pg_{\mu\nu}.
\end{equation}

For the FLRW universe with metric
\begin{equation}
ds^2=-N^2dt^2+a(t)^2\left(\frac{dr^2}{1-kr^2}+r^2d\Omega^2\right),
\end{equation}
where $d\Omega^2$ is the metric in the unit sphere and $k=-1,0,1$ for the open, flat and closed universe, respectively, the four velocity of the fluid is given by $U_{\mu}=N\delta^{0}_{\mu}$ so that  
\begin{equation}
\mu=(\dot{\phi}+\theta\dot{S})/N.
\end{equation}
On the other hand, by thermodynamical considerations, Lapchinskii and Rubakov \cite{lapchinskii} found that the expression for the pressure is given in terms of the potentials by
\begin{equation}\begin{aligned}
p&=\frac{w\mu^{1+1/w}}{(1+w)^{1+1/w}}e^{-S/w}\\&=\frac{w}{(1+w)^{1+1/w}}\left(\frac{\dot{\phi}+\theta\dot{S}}{N}\right)^{1+1/w}e^{-S/w}.
\end{aligned}\end{equation}
For the particular case of FLRW universes, we have
\begin{equation}
K_{ij}=\frac{\dot{a}}{Na}h_{ij},\,\,\, R_{ij}=\frac{2k}{a^2}h_{ij},
\end{equation}
with $h_{ij}=\text{diag}\left(\frac{1}{1-kr^2},r^2,r^2\sin^2{\theta}\right)$, so that the total action is given by (in units where $16\pi G=1$)
\begin{equation}\begin{aligned}
S&=\int{dtd^3{\bf x}N\sqrt{h}\left(K_{ij}K^{ij}-K^2+R\right)}+\int{dtd^3{\bf x}N\sqrt{h}p}\\
&=\int \frac{r^2 \sin{\theta}}{\sqrt{1-kr^2}}d^3x\int dt\Bigg\{-6\frac{\dot{a}^2a}{N}+6kNa\\&+N^{-1/w}a^3\frac{w}{(1+w)^{1+1/w}}(\dot{\phi}+\theta\dot{S})^{1+1/w}e^{-S/w}\Bigg\}.
\end{aligned}\end{equation}
The spatial integration  does not affect the equations of motion, and we have the following canonical momenta associated, respectively, to the dynamical variables $a$, $\phi$ and $S$
\begin{equation}\begin{aligned}
&p_a=-\frac{12\dot{a}a}{N},\,\,\,p_\phi=\frac{N^{-1/w}a^3}{(1+w)^{1/w}}(\dot{\phi}+\theta\dot{S})^{1/w}e^{-S/w},\,\,\,\\&p_S=\theta p_\phi.
\end{aligned}\end{equation}
The Hamiltonian of the system will be given by
\begin{equation}
H=p_a\dot{a}+p_{\phi}(\dot{\phi}+\theta\dot{S})-L,
\end{equation}
where
\begin{equation}\begin{aligned}
L=&-6\frac{\dot{a}a}{N}+6kNa+N^{-1/w}a^3\frac{w}{(1+w)^{1+1/w}}\times\\&\times(\dot{\phi}+\theta\dot{S})^{1+1/w}e^{-S/w}.
\end{aligned}\end{equation}
After a tedious but straightforward calculation, we find
\begin{equation}
H=N\left(-\frac{p_a^2}{24a}-6ka+p_{\phi}^{1+w}a^{-3w}e^{-S/w}\right).
\end{equation}

Since the action does not depend on $\dot{N}$, we conclude that $N$ is actually a Lagrange multiplier of the theory. This is not surprising since the results could not depend on how the spacetime is sliced. Varying the action
\begin{equation}
S=\int{dt \left[p_a\dot{a}+p_{\phi}(\dot{\phi}+\theta\dot{S})-H\right]}
\end{equation}
with respect to $N$ leads to the super-Hamiltonian constraint
\begin{equation}
\mathcal{H}=-\frac{p_a^2}{24a}-6ka+p_{\phi}^{1+w}a^{-3w}e^{-S/w}\approx 0.
\end{equation}
Performing a canonical transformation of the form
\begin{equation}\begin{aligned}
&T=-p_Se^Sp_\phi^{-(1+w)},\,\,\,p_T=p_\phi^{(1+w)}e^S,\\&\bar{\phi}=\phi+(1+w)\frac{p_S}{p_\phi},\,\,\,\bar{p}_\phi=p_\phi,
\end{aligned}\end{equation}
we get
\begin{equation}
\mathcal{H}=-\frac{p_a^2}{24a}-6ka+\frac{p_T}{a^{3w}}\approx 0.
\end{equation}
Now, we proceed with Dirac's algorithm of quantization of constrained systems by making the substitutions $p_a\to-i\partial/\partial a$, $p_T=-i\partial/\partial T$, and demand that the constraint annihilate the wave function, finding the Schr\"odinger-Wheeler-DeWitt equation of the universe
\begin{equation}
\frac{\partial^2\Psi}{\partial a^2}+144ka^2\Psi+i24a^{1-3w}\frac{\partial \Psi}{\partial t}=0,
\end{equation}
with $t=-T$ being the time coordinate in the gauge $N=a^{3w}$, as follows from Hamilton's classical equations of motion \cite{alvarenga2}.
Notice that the above equation is of the form $i\partial \Psi/\partial t=\hat{H}\Psi$. In order to the Hamiltonian operator $\hat{H}$ to be self-adjoint we define the internal product of two wave functions as 
\begin{equation}
\left<\Phi,\Psi\right>=\int_{0}^{\infty}{a^{1-3w} \Phi^{\ast}\Psi da}
\label{internal product}
\end{equation}
and impose restrictive boundary conditions at $a=0$. The simplest ones are the Dirichlet and Neumann conditions:
\begin{equation}\begin{aligned}
&\Psi(0,t)=0,\,\,\,\text{(Dirichlet)}\\
&\frac{\partial \Psi(0,t)}{\partial a}=0,\,\,\, \text{(Neumann)}.
\end{aligned}\end{equation}
As we will see, the situation is qualitatively different in the HL theory of gravity.

\section{Quantum Cosmology in HL}

The total action we will consider here is
\begin{equation}
S=S_{HL}+\int{dtd^3{\bf x}N\sqrt{h}p},
\end{equation}
where $S_{HL}$ is given by Eq. (\ref{HL action}). We choose this action basically 
because the GR action can be recovered in the low-energy limit.
Discarding the spatial integration again, we have
\begin{equation}\begin{aligned}
S=&\int dt\Bigg[-3(3\lambda-1)\frac{\dot{a}^2a}{N}+6Nka-2\Lambda a^3\\&-\frac{12kN}{a}(3g_2+g_3)-\frac{24kN}{a^3}(9g_4+3g_5+g_6)\\&+N^{-1/w}a^3\frac{w}{(1+w)^{1+1/w}}(\dot{\phi}+\theta\dot{S})^{1+1/w}e^{-S/w}\Bigg]
\label{action}
\end{aligned}\end{equation}
Let us introduce the following constants (as in Ref. \cite{bertolami}):
\begin{equation}\begin{aligned}
&g_C=6k, \,\,\, g_{\Lambda}=2\Lambda, g_{r}=12k(3g_2+g_3),\\& g_S=24k(9g_4+3g_5+g_6).
\end{aligned}\end{equation}
Now, proceeding as in GR, {\em i.e.}, defining the momenta corresponding to each one of the dynamical variables and calculating the canonical Hamiltonian, we arrive at
\begin{equation}
\mathcal{H}=-\frac{p_a^2}{12(3\lambda-1)a}-g_Ca+g_\Lambda a^3+\frac{g_r}{a}+\frac{g_s}{a^3}+\frac{p_T}{a^{3w}}\approx 0.
\end{equation}

Specializing to the radiation case ($w=1/3$), we find the Schr\"odinger-Wheeler-DeWitt equation
\begin{equation}
\frac{\partial^2\Psi}{\partial a^2}-12(3\lambda-1)\left(g_Ca^2-g_\Lambda a^3-g_r-\frac{g_s}{a^2}\right)\Psi+12(3\lambda-1)i\frac{\partial \Psi}{\partial t}=0,
\label{main equation}
\end{equation}
again with $t=-T$. Note that $g_r$ just shifts the energy levels, since it is a constant in the potential. However the term $g_s$ changes dramatically the effective potential. The case $g_s=0$ corresponds to the detailed balance condition (see \cite{calgani} where, with the use of the detailed balance condition, Calcagni obtain an action similar to Eq. (\ref{action}), but without the term proportional to $a^{-3}$). 

\section{Exact solutions}

\subsection{Flat FLRW universe }
First, note that if we take a spatially flat universe ($k=0$) with $\Lambda=0$,
 we have the following equation  
\begin{equation}
-\frac{1}{12(3\lambda-1)}\frac{\partial^2\Psi}{\partial a^2}=i\frac{\partial \Psi}{\partial t},
\label{free particle}
\end{equation}
which  is a Schr\"odinger-like equation for a free particle with $\hbar=1$ and mass $m_{\lambda}=6(3\lambda-1)$, except for the requirement $a>0$. Let us consider only the case $\lambda>1/3$. In order to ensure the self-adjointness of the above equation, a boundary condition has to be chosen, as discussed in Section III. For sake of simplicity, we choose Dirichlet boundary condition. For an initial wave packet of the form
\begin{equation}
\Psi(a,0)=\left(\frac{128\sigma^3}{\pi}\right)^{1/4}a e^{-\sigma a^2},
\label{normalized initial wave packet}
\end{equation}
Eq. (\ref{free particle}) can be easily solved using a specific propagator (see \cite{lemos}). The result is
\begin{equation}\begin{aligned}
\Psi(a,t)&=\left(\frac{m_\lambda}{m_\lambda+2it\sigma}\right)^{3/2}\left(\frac{128\sigma^3}{\pi}\right)^{1/4}a \\&\times\exp{\left(-\frac{-\sigma m_\lambda^2 a^2}{m_\lambda^2+4\sigma^2t^2}\right)}\exp{\left(\frac{2itm_\lambda\sigma^2a^2}{m_\lambda^2+4\sigma^2t^2}\right)}
\end{aligned}\end{equation}
We can calculate the expectation value of the operator $a$ through the formula
\begin{equation}\begin{aligned}
\left<a\right>(t)&=\left<\Psi,a\Psi\right>=\int_{0}^{\infty}{a|\Psi(a,t)|^2}\\
&=\frac{2}{m_{\lambda}}\sqrt{\frac{2}{\pi\sigma}}\sqrt{\frac{m_\lambda^2}{4}+\sigma^2t^2}.
\end{aligned}\end{equation}

Note that if we take $\lambda=1$, we recover the result obtained in Ref. \cite{lemos}. Nothing changes in HL theory in a flat FLRW universe in comparison with GR (as in the classical case, see for instance \cite{wang}). In particular, $\left<a\right>(t)$ is nonsingular and $\left<a\right>(t)\sim t$ as $t\to \infty$, recovering the classical behavior of the universe in the classical limit. Besides, the evolution of the wave packet is given once we choose a particular boundary condition at $a=0$. Therefore the evolution of the universe is not unique in the sense stated in Sec. II.

\subsection{Closed FLRW universe }

For the spatially closed ($k=1$) FLRW spacetime, there is an effective potential in the Schr\"odinger-Wheeler-DeWitt equation of the universe. From Eq. (\ref{main equation}),
we see that 
 this potential represents a shifted quantum harmonic oscillator with a singular perturbation. Setting the mass  and frequency of the harmonic oscillator in order to have  $m_\lambda=6(3\lambda-1)$ and $\omega_\lambda=\sqrt{2/(3\lambda-1)}$, respectively, we have the following Schr\"odinger-like equation
\begin{equation}
-\frac{1}{2m_\lambda}\frac{\partial^2\Psi}{\partial a^2}+\left(\frac{1}{2}m_\lambda \omega_\lambda^2a^2-g_r-\frac{g_s}{a^2}\right)\Psi=i\frac{\partial \Psi}{\partial t}.
\label{without balance}
\end{equation}
First of all, we will analyze the necessity of a boundary condition at $a=0$ on this equation. The first step is to separate variables in the form $\Psi(a,t)=\psi(a)e^{-i E t}$, leading to
\begin{equation}
\left[-\frac{d^2}{da^2}+V(a)\right]\psi(a)=2m_\lambda E\psi(a),
\end{equation}
with
\begin{equation}
V(a)=m_\lambda^2\omega_\lambda^2a^2-2m_\lambda g_r-(2m_\lambda g_s)/a^2.
\end{equation}
Following Ref. \cite{reed}, we say that $V(a)$ is in the limit circle case at infinity and 
at  zero  if  for all $\lambda$, all solutions of 
\begin{equation}
\left[-\frac{d^2}{da^2}+V(a)\right]\psi(a)=\lambda \psi(a)
\end{equation}
are square integrable at infinity and at zero, respectively. If $V(a)$ is not in the limit circle case, it is said to be in the limit point case.
We now enunciate the Theorem X.7 from Ref. \cite{reed}, which gives us a criterion to decide if the Hamiltonian operator $\hat{H}=-d^2/da^2+V(a)$ is essentially self-adjoint, {\em i.e.}, if it has a unique self-adjoint extension.
\begin{theorem}
Let $V(a)$ be a continuous real-valued function on $(0,\infty)$. Then $\hat{H}=-d^2/da^2+V(a)$ is essentially self-ajoint if and only if $V(a)$ is in the limit point case at both zero and infinity.
\end{theorem}
We need now a criterion to decide if $\hat{H}$ is in the limit point or limit circle case at zero and infinity. We will find this criterion in the next two theorems, extracted again from Ref. \cite{reed}.
\begin{theorem}
Let $V(a)$ be a continuous real-valued function on $(0,\infty)$ and suppose that there exists a positive differentiable function $M(a)$ so that
\begin{itemize}
\item[i)] $V(a)\geq -M(a)$.
\item[ii)] $\int_1^\infty{\sqrt{M(a)}}da=\infty.$
\item[iii)] $M(a)/(M(a))^{3/2}$is bounded near $\infty$.
\end{itemize}
Then $V(a)$ is in the limit point case at $\infty$.
\label{two}
\end{theorem}
\begin{theorem}
Let $V(a)$ be continuous and positive near $a=0$. If $V(a)\geq \frac{3}{4a^2}$ near zero then $-d^2/da^2+V(a)$ is in the limit point case at zero. If for some $\epsilon>0$, $V(a)\leq (\frac{3}{4}-\epsilon)a^{-2}$ near zero, then $-d^2/da^2+V(a)$ is in the limit circle case.
\label{three}
\end{theorem}
From now on, we will consider $g_s< 0$. In the end of this section,  we will return to the case $g_s>0$. Let us first use Theorem \ref{two} to show that the Hamiltonian operator in Eq. (\ref{without balance}) is in the limit point case at infinity. To verify this fact, note that the potential $V(a)$ has a minimum $V_{\text{min}}=2m_\lambda\left[\sqrt{-2g_s}m_\lambda^{1/2}\omega_\lambda-g_r\right]$ at $a=\left(\frac{-2gs}{m_\lambda}\right)^{1/4}\omega_\lambda^{-1/2}$. If $V_{\text{min}}\geq0$ we choose $M(a)=1$ and all the requirements of Theorem \ref{two} are fulfilled. If $V_{\text{min}}\leq 0$, we take $M(a)=\left|V_{\text{min}}\right|$. In any case, we conclude that $\hat{H}$ is in the limit point case at infinity.

Note now that the potential $V(a)$ has the form $V(a)\sim -2m_\lambda g_s/a^2$ near $a=0$.  Therefore, by Theorem \ref{three}, if $-2m_\lambda g_s\geq 3/4$ then  $\hat{H}$ is in the limit point case at zero, otherwise it is in the limit circle case. We have established a range of parameter in which the operator $\hat{H}$ is essentially self-adjoint, {\em i.e.}, if
\begin{equation}
-m_\lambda g_s\geq3/8,
\end{equation}
then the evolution of the wave function representing the universe is uniquely determined by the initial wave packet and no boundary condition at $a=0$ is indeed necessary. Otherwise, we need to impose a boundary condition at this point. For simplicity we did not consider the case $\Lambda\neq 0$, but the previous analysis still works in this case.

It turns out that  Eq. (\ref{without balance}) can be exactly solved. By introducing the new variable $x=\sqrt{m_\lambda\omega_\lambda}a$ and a parameter \cite{bertolami}
\begin{equation}
\alpha=\frac{1}{2}\sqrt{1-8m_\lambda g_s},
\end{equation}
  Eq. (\ref{without balance}) becomes  
\begin{equation}
-\frac{d^2}{dx^2}\psi(x)+\left[x^2-\frac{2}{\omega_\lambda}(gr+E)+\frac{4\alpha^2-1}{4x^2}\right]\psi(x)=0.
\end{equation}
By introducing the new function $y(\eta)$ given by
\begin{equation}
\psi(x)=e^{-\frac{x^2}{2}}x^{\alpha+1/2}y(x^2),
\end{equation}
it is easy to see that $y(\eta)$ satisfies the associated Laguerre equation
\begin{equation}
\eta y''(\eta)+(1+\alpha-\eta)y'(\eta)+\lambda_Ey(\eta)=0,
\label{laguerre}
\end{equation}
where 
\begin{equation}
\lambda_E=\left[\frac{(E+g_r)}{2\omega_\lambda}-\frac{(1+\alpha)}{2}\right].
\end{equation}
This equation is known to be Hermitian (formally self-adjoint) with the inner product
\begin{equation}
\left<f,g\right>_L=\int_{0}^{\infty}{e^{-\eta}\eta^{\alpha}\overline{f(\eta)}g(\eta)d\eta}.
\end{equation}
The general solution of Eq. (\ref{laguerre}) is given by \cite{abramowitz}
\begin{equation}
y(\eta)=A_EM(-\lambda_E,1+\alpha,\eta)+B_EU(-\lambda_E,1+\alpha,\eta),
\end{equation}
where $M$ and $U$ are the confluent hypergeometric functions of first and second kinds, respectively. For $\lambda_E=n=0,1,2,\dots$, both $M$ and $U$ are polynomials of degree $n$, proportional to the associated Laguerre polynomial $L_n^\alpha(\eta)$. The other linear independent solution is not square-integrable near $a=0$, so it must be excluded from the
present analysis.
If $\lambda_E\notin \mathbb{N}\cup \{0\}$, we have the following asymptotic behavior for $M$ and $U$, as $\eta\to\infty$:
\begin{equation}\begin{aligned}
&M(-\lambda_E,1+\alpha,\eta)\sim \frac{\Gamma(1+\alpha)}{\Gamma(-\lambda_E)}e^{\eta}\eta^{-1-\lambda_E-\alpha},\\
&U(-\lambda_E,1+\alpha,\eta)\sim \eta^{\lambda_E}.
\end{aligned}\end{equation}
Therefore $M$ is not square-integrable near infinity, meanwhile $U$ is. So $M$ is not an acceptable solution in this case. 
As $\eta\to0$, the asymptotic behavior of $U$ is given by
\begin{equation}
U(-\lambda_E,1+\alpha,\eta)\sim \eta^{-\alpha}\Gamma(\alpha)/\Gamma(-\lambda_E),
\label{asymptotic}
\end{equation}
and, hence, $U$ is square-integrable near $a=0$ only if $\alpha<1$. For $\alpha\geq1$, we do not have an acceptable solution, except in the case $\lambda_E=n=0,1,2,\dots$. Therefore, for $\alpha\geq 1$ (which corresponds to $-m_\lambda g_s\geq 3/8$), we quantize automatically the energy levels of the universe, which are given by
\begin{equation}
E_n=(2n+1+\alpha)\omega_\lambda-g_r.
\end{equation}
The corresponding normalized eigenstates are given by
\begin{equation}\begin{aligned}
\Psi_n(a,t)=&(4m_\lambda \omega_\lambda)^{1/4}\left[\frac{\Gamma(n+1)}{\Gamma(\alpha+n+1)}\right]^{1/2}(m_\lambda \omega_\lambda a^2)^{\frac{2\alpha+1}{4}}\\&\times\exp{\left(-\frac{m_\lambda \omega_\lambda}{2}a^2\right)}L_n^\alpha(m_\lambda \omega_\lambda a^2)e^{-i E_n t},
\label{eingenfunctions}
\end{aligned}\end{equation}
and a general solution $\Psi(a,t)$, depending on the initial wave packet $\Psi(a,0)$, is then given by
\begin{equation}
\Psi(a,t)=\sum_{n=0}^{\infty}{c_n\Psi_n(a,t)},
\label{sum}
\end{equation}
with
\begin{equation}
c_n=\int_{0}^{\infty}{\Psi(a,0)\Psi_n^{\ast}(a,0)da}.
\end{equation}

For $1/2\leq\alpha<1$, Eq. (\ref{asymptotic}) shows that $U$ is square-integrable at $a=0$. We then need a boundary condition at $a=0$ in order to have a well posed Sturm-Liouville problem. They are found in \cite{derkach} and are given by
\begin{equation}
\Gamma_1y=\theta \Gamma_0y,\,\,\,\,\, \theta \in \mathcal{R},
\end{equation}
where
\begin{equation}\begin{aligned}
&\Gamma_0y=\lim_{x\to 0}{x^{\alpha+1}y'(x)},\\
&\Gamma_1y=\lim_{x\to 0}{\left[y(x)+\frac{x}{\alpha}y'(x)\right]}.
\end{aligned}\end{equation}
If $\theta=\infty$, then $\lim_{x\to 0}{x^{\alpha+1}y'(x)}=0$ and we find \cite{everitt} $\lambda_E=n=0,1,2,\dots$ and the corresponding associated Laguerre polynomials. For others values of $\theta$, the quantized energy levels are not so simple. In all cases, the wave function will satisfy the DeWitt condition
\begin{equation}
\Psi(0,t)=0,
\end{equation}
but we stress the fact that this is not the condition which turns the Wheeler-DeWitt equation into a self-adjoint form.

It is worth to analyze the case $\lambda\to1$, $g_r\to0$ and $\alpha\to 1/2$, where we expect to recover the usual quantum cosmology. In this limit, Eq. (\ref{eingenfunctions}) becomes
\begin{equation}\begin{aligned}
\Psi_n(a,t)=&\left(\frac{\sqrt{48}}{\sqrt{\pi}2^{2n+1}(2n+1)!}\right)^{1/2}H_{2n+1}(\sqrt{12}a)\\&\times e^{-6a^2}e^{-i E_n t},
\label{eingenfunctions1}
\end{aligned}\end{equation}
with
\begin{equation}
E_n=2n+3/2.
\end{equation}
These same eigenstates have been already found in \cite{lemos}, in the context of ordinary quantum cosmology. They satisfy the Dirichlet boundary condition $\Psi(0,t)=0$. Neumann boundary condition is satisfied when $\alpha\to-1/2$. We do not consider this case here, but it is trivial to generalize our results to $-1<\alpha<1/2$ (see \cite{derkach}). Therefore, the   HL quantum cosmology tend naturally to usual quantum cosmology in the appropriate limit.

Have studied the self-adjointness of the evolution equation of the universe, let us now focus on the evolution of the expectation value of the scale factor given a solution representing the state of the universe. Obviously, if we calculate the expectation value of the scale factor in any of these eigenstates, it will be constant. But the universe evolves so that we must consider wave packets representing the state of the universe. In order to find exact solutions, we choose an initial wave packet of the form
\begin{equation}
\Psi(a,0)=\left[\frac{2^{\nu+5/2}\sigma^{\nu+3/2}}{\Gamma\left(\nu+3/2\right)}\right]^{1/2}a^{\nu+1}e^{-\sigma a^2},
\label{initial nu}
\end{equation}
where 
$g_s=-\nu(\nu+1)/(2m_\lambda)$. Note that $g_s< 0$ in this case, so that the potential $V(a)$ is repulsive, preventing the formation of a classical singularity.
The propagator for Eq. (\ref{without balance}) is given by \cite{efthimiou}
\begin{equation}\begin{aligned}
G(a,a';t;\nu)=&\frac{m_\lambda\omega_\lambda\sqrt{aa'}}{\sin{(\omega_\lambda t)}}e^{ig_r t}i^{-(\nu+3/2)}\\&\times\exp{\left[\frac{im_\lambda\omega_\lambda}{2}\cot{(\omega_\lambda t)(a^2+a'^2)}\right]}\\&\times J_{\nu+1/2}\left(\frac{m_\lambda\omega_\lambda a a'}{\sin{(\omega_\lambda t)}}\right),
\end{aligned}\end{equation}
and through the equation
\begin{equation}
\Psi(a,t)=\int_{0}^{\infty}{G(a,a';t;\nu)\Psi(a',0)da'},
\end{equation}
we find, after some tedious calculation,
\begin{equation}\begin{aligned}
\Psi(a,t)=&\left[\frac{2^{\nu+5/2}\sigma^{\nu+3/2}}{\Gamma\left(\nu+3/2\right)}\right]^{1/2}\left(\frac{m_\lambda \omega_\lambda}{i \sin{(\omega_\lambda t)}}\right)^{\nu+3/2}\\&\times\frac{a^{\nu+1}e^{ig_r t}}{\left[2\sigma-im_\lambda\omega_\lambda \cot{(\omega_\lambda t)}\right]^{\nu+3/2}}\\&\times\exp{\left(\frac{-m_\lambda^2\omega_\lambda^2\sigma a^2}{4\sigma^2\sin^2{(\omega_\lambda t)}+m_\lambda^2\omega_\lambda^2\cos^2{(\omega_\lambda t)}}\right)}\\&\times\exp{\Bigg[\frac{im_\lambda\omega_\lambda\sin{(\omega_\lambda t)}\cos{(\omega_\lambda t)}a^2}{2(4\sigma^2\sin^2{(\omega_\lambda t)}+m_\lambda^2\omega_\lambda^2\cos^2{(\omega_\lambda t)})}}\\&(4\sigma^2-m_\lambda^2\omega_\lambda^2)\Bigg].
\end{aligned}\end{equation}
The expectation value of the scale factor is given by
\begin{equation}\begin{aligned}
\left<a\right>(t)&=\frac{\Gamma(\nu+2)}{\sqrt{2\sigma}\Gamma\left(\nu+3/2\right)}\frac{1}{m_{\lambda}\omega_\lambda}\\&\times\sqrt{4\sigma^2\sin^2{(\omega_\lambda t)}+m_{\lambda}^2\omega_{\lambda}^2\cos^2{(\omega_\lambda t)}}\\&=\frac{\left<a\right>(0)}{m_\lambda\omega_\lambda}\sqrt{4\sigma^2\sin^2{(\omega_\lambda t)}+m_{\lambda}^2\omega_{\lambda}^2\cos^2{(\omega_\lambda t)}}.
\end{aligned}\end{equation}  
We stress the nonsingular character of the above equation, since $\left<a\right>(t)\neq 0$ for all times. As  in the usual quantum cosmology \cite{lemos}, the singularity is not present in the quantum model. Note also the similarity with the results found in \cite{lemos}. The only difference here is that the parameter $\lambda$ changes the frequency of oscillation of the scale factor. The behavior of this quantum universe in HL theory is not much different from usual quantum cosmology.

\begin{figure}[!htb]
\centering
\includegraphics[scale=0.6]{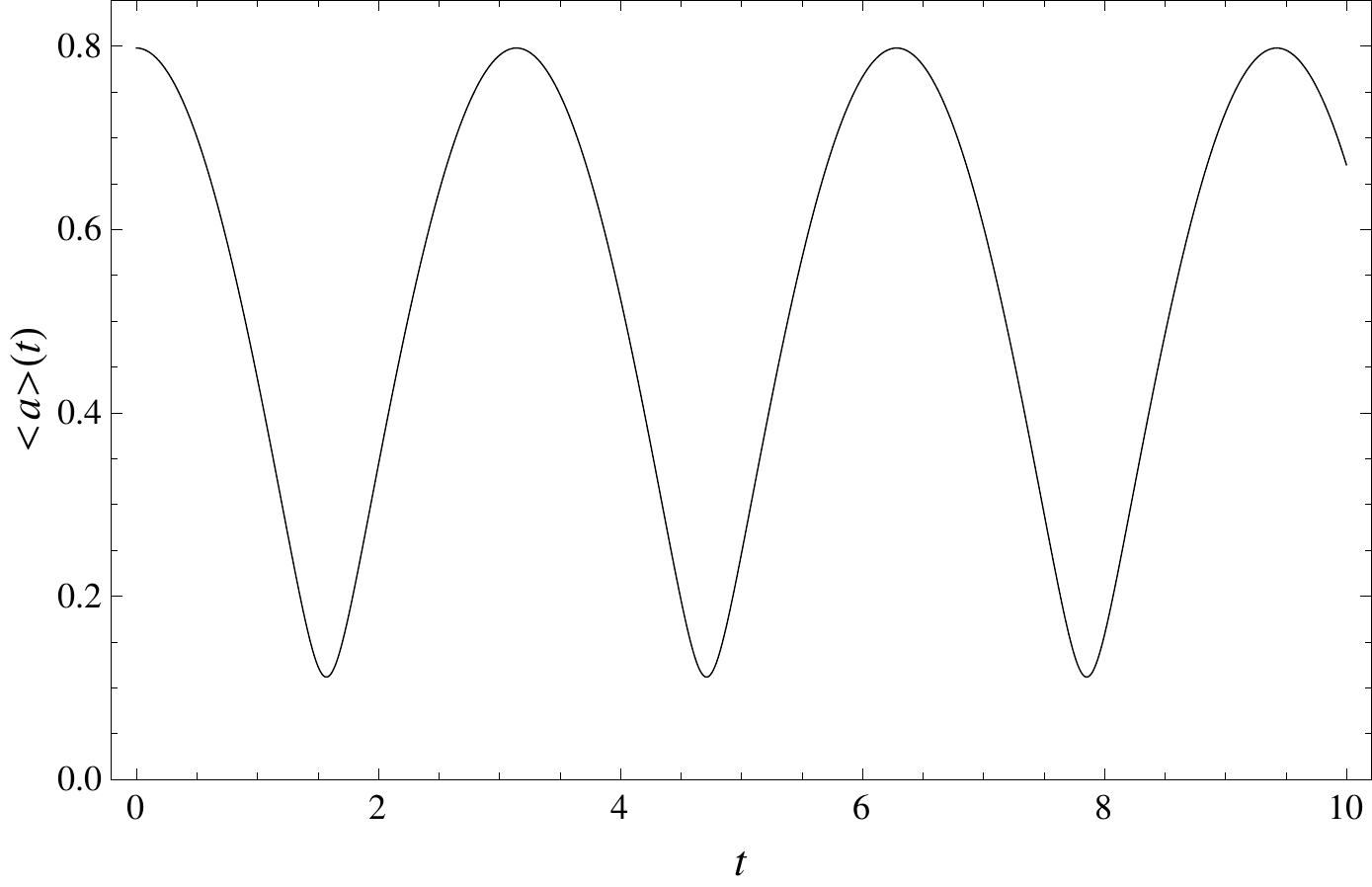}
\caption{The evolution of the scale factor in the case of an attractive potential. Twenty terms in expansion (\ref{sum}) were used in this numerical approximation.}
\label{fig}
\end{figure}

We can also study the case $g_s>  0$, when the potential $V(a)$ is attractive, not preventing the formation of a classical singularity. In order to obtain exact solutions we must restrict $2m_\lambda g_s\leq 1/8$ so that $\alpha\geq 0$. We do not have a propagator in this case, but we can use expression (\ref{sum}) to find numerically the evolution of the scale factor. This in done in Fig. \ref{fig}, where we have chosen an initial wave packet of the form
\begin{equation}
\Psi(a,0)=2 \left(\frac{8}{\pi}\right)^{1/4}ae^{-a^2}.
\end{equation}
The parameters characterizing HL gravity are $\lambda=1$, $gr=0$ and $\alpha=1/4$. Note that $\left<a\right>(t)\neq 0$ $\forall$ $t$. The singularity has been excluded.
\section{Concluding Remarks}

We have seen that in the HL theory of gravity, it is possible not only to exclude the initial big bang singularity, but also to determine uniquely the evolution of the wave function of the universe given an initial wave packet. An equivalent statement is that no boundary condition at $a=0$ is necessary in a quantum cosmology in the context of HL gravity. In general, theories
of gravity do not tell us which boundary condition we must choose, so it is a remarkable fact that one of these theories excludes this ambiguity. Moreover, in HL quantum cosmology, the evolution of the expectation value of the scale factor resembles the evolution found in usual quantum cosmology, the only difference being the frequency of oscillation of the bouncing universe. It is interesting to notice that the quantum regime of the HL theory of gravity can also provide  a
viable framework for the description of the ``asymptotic darkness'' of the visible
universe \cite{Elizalde:2011tx}. Our early universe results are in a certain sense
 complementary to the asymptotic regime described in \cite{Elizalde:2011tx}. It is certainly worthy to further explore  the connections between the two approaches.

\acknowledgements 

The authors thank FAPESP and CNPq for the financial support.


\end{document}